\input harvmac
\newcount\figno
\figno=0
\def\fig#1#2#3{
\par\begingroup\parindent=0pt\leftskip=1cm\rightskip=1cm\parindent=0pt
\baselineskip=11pt
\global\advance\figno by 1
\midinsert
\epsfxsize=#3
\centerline{\epsfbox{#2}}
\vskip 12pt
{\bf Fig. \the\figno:} #1\par
\endinsert\endgroup\par
}
\def\figlabel#1{\xdef#1{\the\figno}}
\def\encadremath#1{\vbox{\hrule\hbox{\vrule\kern8pt\vbox{\kern8pt
\hbox{$\displaystyle #1$}\kern8pt}
\kern8pt\vrule}\hrule}}

\overfullrule=0pt

%
\def\tilde{\widetilde}
\def\cqg#1#2#3{{\it Class. Quantum Grav.} {\bf #1} (#2) #3}
\def\np#1#2#3{{\it Nucl. Phys.} {\bf B#1} (#2) #3}
\def\pl#1#2#3{{\it Phys. Lett. }{\bf #1B} (#2) #3}
\def\prl#1#2#3{{\it Phys. Rev. Lett.}{\bf #1} (#2) #3}
\def\physrev#1#2#3{{\it Phys. Rev.} {\bf D#1} (#2) #3}

\def\cmp#1#2#3{{\it Comm. Math. Phys.} {\bf #1} (#2) #3}
\def\mpl#1#2#3{{\it Mod. Phys. Lett. }{\bf #1} (#2) #3}
\def\ijmp#1#2#3{{\it Int. J. Mod. Phys.} {\bf #1} (#2) #3}

\font\zfont = cmss10 

\def\bigone{\hbox{1\kern -.23em {\rm l}}}
\def\ZZ{\hbox{\zfont Z\kern-.4emZ}}

\def\a{\alpha}
\def\b{\beta}
\def\g{\gamma}
\def\d{\delta}
\def\e{\epsilon}

\def\m{\mu}
\def\n{\nu}
\def\x{\xi}
\def\r{\rho}

\def\G{\Gamma}
\def\D{\Delta}

\def\O{\Omega}
\def\o{\over}

\def\tria{$\triangleright $}
\def\no{\noindent}

\Title{NSF-ITP-96-19, hep-th/9605053,}
{\vbox{\centerline{${\cal M}$-Theory on Eight-Manifolds}}}
\smallskip
\centerline{Katrin Becker}
\smallskip
\centerline{\it Institute for Theoretical Physics, University of California}
\centerline{\it Santa Barbara, CA 93106-4030}
\smallskip
\centerline{and}
\smallskip
\centerline{Melanie Becker}
\smallskip
\centerline{\it Department of Physics, University of California }
\centerline{\it Santa Barbara, CA 93106-9530}\bigskip
\baselineskip 18pt
\vskip 1cm
\noindent
We show that in certain compactifications of ${\cal M}$-theory 
on eight-manifolds to three-dimensional Minkowski space-time 
the four-form field strength can have a non-vanishing expectation value,
while an $N=2$ supersymmetry is preserved. 
For these compactifications
a warp factor for the metric has to be taken into account. This warp factor 
is non-trivial in three space-time dimensions due to 
Chern-Simons corrections
to the fivebrane Bianchi identity.
While the original metric on the internal space is not K\"ahler, 
it can be conformally transformed to a metric that is K\"ahler and
Ricci flat, so that the internal manifold has $SU(4)$ holonomy.

\Date{May, 1996}
\newsec{Introduction}
The duality symmetries between different string theories 
can be naturally understood
from ${\cal M}$-theory 
\ref\mt{E.~Witten, ``String Theory Dynamics in Various Dimensions'',
\np {443}{1995}{85}, hep-th/9503124.}\ref\hulltow{C.~M.~Hull and 
P.~K.~Townsend, ``Enhanced Gauge Symmetries in
Superstring Theories, \np {451}{1995}{525}, hep-th/9505073.}
or its twelve-dimensional generalization, that has been 
called ${\cal F}$-theory
\ref\vafa{C.~Vafa, ``Evidence for F-Theory'', preprint HUTP-96-A004, 
hep-th/ 9602022.}
\ref\maku{D.~Kutasov and E.~Martinec, ``New Principles for String/Membrane
Unification'', preprint EFI-96-04, hep-th/9502049.}.
${\cal M}$-theory contains membranes and fivebranes, which 
turn out to be dual in
eleven dimensions.
Membrane-fivebrane duality predicts the existence of a space-time 
correction to the 
eleven-dimensional supermembrane action
\ref\dlm{M.~J.~Duff, J.~T.~Liu and R.~Minasian,
``Eleven Dimensional Origin of String/String Duality: A One Loop Test'', 
\np {452} {1995} {261}.}.
Taking this anomaly into account, it is our goal in this paper to examine 
the conditions under which the ground state of ${\cal M}$-theory can
 be supersymmetric and 
of the form $M^3\times K^8$, where $M^3$ is three-dimensional 
Minkowski space-time
and $K^8$ is the internal eight-manifold. Compactifications of 
$\cal M$-theory
\ref\ef{E.~Witten, ``Nonperturbative Superpotentials in String Theory'', 
preprint IASSNS-HEP-96-29, hep-th/9604030.}
and $\cal F$-theory {\vafa} on eight-manifolds are fascinating, since they
may lead us to a way to understand the dynamics of $N=1$ supersymmetric
field theories and string theories in $D=4$, supersymmetry breaking and
 to the solution of the cosmological constant problem along the lines 
proposed by Witten 
\ref\edcc{E.~Witten, ``Is Supersymmetry Really Broken?'', \ijmp {A10} {1995} 
{1247}, hep-th/9409111; ``Strong Coupling and the Cosmological Constant'', 
\mpl {A10} {1995} {2153}, hep-th/9506101.}\foot{For a field theory 
example see
\ref\bbscc{K.~Becker, M.~Becker and A.~Strominger, ``Three-Dimensional 
Supergravity and the Cosmological Constant'', \physrev{51} {1195} {6603}, 
hep-th/9502107.}.}.
Our computation shows the existence of new vacua of $\cal M$-theory
having $N=2$ supersymmetry for which the four-form field 
strength can have a non-vanishing expectation 
value, while the three-dimensional cosmological constant vanishes.

\newsec{${\cal M}$-Theory on Eight-Manifolds}
The bosonic part of the 
action of the eleven-dimensional supergravity limit of ${\cal M}$-theory 
is given by
 \ref\cjs{E.~J.~Cremmer, B.~Julia and J.~Scherk, 
``Supergravity Theory in 11 Dimensions'', \pl {5} {1978} {409}.}:
\eqn\ai{ {\cal S}_{11}={1\o 2} \int d^{11} x \sqrt{-{\hat g} }\left[
{\hat R} -{1\o 2} \hat F_4 \wedge * \hat F_4  
-{1\o 6 } \hat C_3 \wedge \hat F_4 \wedge \hat F_4 \right]   ,  }
where ${\hat g}_{MN} $ is the space-time metric (the hat 
denotes eleven-dimensional quantities) and $\hat C_3$ 
is a three-form with field strength $\hat F_4=d\hat C_3 $. We have set the
gravitational constant 
equal to one. 
The complete action is invariant under local supersymmetry transformations 
\eqn\aii{
\eqalign{
& \d {\hat e_M}^A=i{\bar \eta} {\hat \G}^A \psi_M,  \cr
& \d \hat C_{MNP}=3 i{\bar \eta} {\hat \G}_{[MN} \psi _{P]}, \cr 
& \d \psi_M ={\hat \nabla} _M \eta - {1\o 288} 
\left( {{\hat \G}_M}^{\,\,\,\,\, PQRS}-8 
{\hat \d}_M^P {\hat \G}^{QRS} \right) \hat F_{PQRS} \eta,\cr} }
where ${\hat e_M}^A$ is the vielbein, $\psi_M$ is the gravitino,   
$\eta$ is an eleven-dimensional anticommuting Majorana spinor
and ${\hat\nabla}_M$ denotes the covariant derivative
involving the Christoffel connection as usual. 
Further notations and conventions will be given in the appendix.
 
The field strength obeys the Bianchi identity: 
\eqn\aaiaai{d\hat F_4=0,}
or in components $\partial_{[M} \hat F_{PQRS]}=0$. This equation is metric 
independent. The field equation for $\hat F_4$ is: 
\eqn\aixxiaa{ d * \hat F_4=-{1\o 2} \hat F_4^2,  }
or in components after dualizing
\eqn\aiv{\hat E^{-1} {\partial}_M ({\hat E} {\hat F}^{MNPQ})-{1 \o 1152}
{\hat \e}^{NPQRSTUVWXY}
 \hat F_{RSTU} \hat F_{VWXY}=0,  }
where ${\hat E}=\det {\hat e_M}^A$. The fivebrane soliton 
appears as a solution to the eleven-dimensional
field equations and it couples to the dual seven-form field strength
$\hat F_7=*\hat F_4$. Equation {\aixxiaa} then becomes 
the Bianchi identity for the eleven-dimensional fivebrane.

This equation has in general gravitational 
Chern-Simons corrections associated to the sigma-model 
anomaly on the six-dimensional fivebrane worldvolume {\dlm}. 
The corrected fivebrane Bianchi identity takes the form
\eqn\aixxi{ d * \hat F_4=-{1\o 2} \hat F_4^2 +(2\pi)^4 { \b} {X_8},  }
where ${\beta}$ is related to the fivebrane tension 
by $T_6=1/ (2\pi)^3 {\beta }$. Henceforth we set 
${\beta}=1$. Since the gauge-fixed 
theory of the fivebrane
is described by a chiral anti-self-dual tensor multiplet, 
the eight-form anomaly polynomial is expressed in terms of the 
Riemann tensor
\ref\agw{L.~Alvarez-Gaum\' e and E.~Witten,``Gravitational Anomalies'', 
\np{234} {1983} {269}.}
\eqn\avii{
{ X_8}={1\o (2\pi)^4}\left (-{1\o 768 } ({\rm tr} {\hat R}^2 )^2 
+{1\o 192} {\rm tr} {\hat R}^4\right).   } 

The anomaly leads to an additional term in the action {\ai}
\eqn\aviii{
\d {\cal S}_{11} ={1\o 2} \int {\hat C}_3 
\wedge \left( -{1\o 768} ({\rm tr}
 {\hat R}^2 )^2 +{1\o 192} 
{\rm tr} 
{\hat R}^4 \right) . }
The existence of this interaction can be verified by computing the 
one-point function of the two-form 
$B_{MN}$ in the type IIA string theory compactified 
on an eight-manifold
\ref\vawi{C.~Vafa and E.~Witten,``A One-Loop Test of 
String Duality'', \np{447} {1995} {261}.}. The result of this 
calculation has no dilaton dependence, 
since this would spoil gauge invariance. It can therefore 
be extrapolated to 
eleven dimensions and it gives the previous answer.

A supersymmetric configuration is one that obeys 
for some Majorana spinor $\eta$ the conditions
\eqn\aix{
\eqalign{
& \d_{\eta} {\hat e_M}^A=0, \cr
& \d_{\eta} \hat C_{MNP}=0, \cr 
& \d_{\eta} \psi_M =0. \cr} }
Since in the background the spinor $\psi_M$ vanishes, 
the first two of the above equations 
are satisfied, and only the gravitino equation remains to be solved
\eqn\aaiixx{
{\hat \nabla} _M \eta - {1\o 288} 
\left( {{\hat \G}_M}^{\,\,\,\,\, PQRS}-8 {\hat \d}_M^P 
{\hat \G}^{QRS} \right)
 \hat F_{PQRS} \eta =0.}
The most general ansatz for the metric that is consistent 
with maximal symmetry is\foot{Compactifications 
of eleven-dimensional supergravity to $D=4$ anti-de Sitter 
space with a warp factor but without the anomaly {\aviii}
have been considered before in 
\ref\wini{B.~de~Wit, H.~Nicolai and N.~P.~Warner, ``The Embedding 
of Gauged $N=8$ Supergravity into $d=11$ Supergravity'', 
\np {255}{1985}{29}.}. In these theories the four-form 
field strength is proportional to the cosmological constant 
of the external space and vanishes therefore for compactifications 
to Minkowski space. Compactifications of 
type II superstring theories with a warp factor have also been 
discussed by \ref\wshd{B.~De Wit, D.~J.~Smit and 
N.~D.~Hari Dass, ``Residual 
Supersymmetry of Compactified $d=10$ Supergravity'', \np {283}{1986}
{165}.}. For the heterotic string the warp factor is necessary 
in order to obtain solutions with torsion 
\ref\strom{A.~Strominger, ``Superstrings with Torsion'', 
\np {274} {1986} {253}.}.}

\eqn\gi{ {\hat g}_{MN}(x,y)=\D(y)^{-1} g_{MN} (x,y), }
where 
\eqn\ax{g_{MN}(x,y) =\left( \matrix{ g_{\m \n}(x)  & 0 \cr 
0 & g_{mn}(y) \cr }\right).  }
Here $x$ are the three-dimensional external coordinates labeled 
by the indices $\m,\n,\dots$ and $y$ the ones of the 
Euclidean eight-manifold labeled by $m,n,\dots$. 
$\Delta(y)$ is a scalar function called the ``warp factor''.
We first would like to rewrite {\aaiixx} in terms of $g_{MN}$. 
We can relate covariant derivatives with respect to conformally 
transformed metrics by using the formula:
\eqn\di{{\hat \nabla }_M\eta
=\nabla_M \eta +{1\o 2} \O^{-1} {\G_M}^N 
(\nabla_N \O) \eta,  }
where ${\hat g}_{MN}=\O^2 g_{MN}$ . This gives the relation 
\eqn\dii{{\hat \nabla}_M \eta 
=\nabla_M \eta -{1\o 4} \D^{-1} 
{\G_M}^N (\nabla_N \D) \eta. }
Furthermore, ${\hat \G}_M$ matrices are related to $\G_M$ matrices as
\eqn\diii{{\hat \G}_M =\D^{-1/2} \G_M \qquad {\rm and} \qquad 
{\hat \G}^M =\D^{1/2} \G^M, }
while ${\hat F}_{MNPQ}$ will be kept fixed under 
the transformation {\gi}. 
We then obtain for  
{\aaiixx}  in terms of $g_{MN}$ the result: 
\eqn\div{ \nabla_M \eta -{1\o 4} {\G_M}^N \partial_N (\log \D )\eta 
-{ 1\o 288} \D^{3/2}\left( {\G_M}^{PQRS} -8 \d_M^P \G^{QRS} \right) 
F_{PQRS} \eta =0.}
We make a decomposition of the gamma matrices that is appropriate to the
$11=3+8$ split, by taking 
\eqn\axi{\eqalign{
&\G_{\m}=  \g_{\mu} \otimes \g_9,  \cr 
& \G_m= 1  \otimes \g_m,  \cr}} 
where $\g_{\mu}$ and $\g_m$ are the gamma matrices of $M^3$ and $K^8$ 
respectively and $\g_9$ is the eight-dimensional 
chirality operator, that satisfies $\g_9^2=1$ and anti-commutes with all
the $\g_m$'s. 
   
We decompose the eleven-dimensional spinor $\eta$ as a sum of 
terms of the form
\eqn\axxii{\eta=\e \otimes \x,}
where $\e$ is a three-dimensional anticommuting spinor, 
while $\x$ is a commuting eight-dimensional Majorana-Weyl spinor.
Spinors of the form {\axxii} that solve $\d_{\eta} \a=0$ for 
every field $\a$, give unbroken 
supersymmetries. We shall be interested in compactifications 
having $N=2$ supersymmetry in three dimensions for which 
two spinors on $K^8$ of the same 
chirality can be found. 
We can combine these real spinors into a complex spinor of a well defined 
chirality. Without loss of generality we will take the 
chirality to be positive.   
Compactifications for which spinors  
of the previous form can be found will, in general, have
$\int X^8 \neq 0$. 

In \ref\ipw{C.~J.~Isham and C.~N.~Pope, 
``Nowhere-vanishing Spinors and Topological Obstruction 
to the Equivalence of the NSR and GS Superstrings'', \cqg {5} {1988} {257}; 
C.~J.~Isham, C.~N.~Pope and N.~P.~Warner, 
``Nowhere-vanishing Spinors and Triality Rotations in 
8-Manifolds'', \cqg{5} {1988} {1297}. }
it was shown that demanding the existence of a nowhere-vanishing 
eight-dimensional
Majorana-Weyl spinor in the $8_c$ representation of $SO(8)$ gives 
a relation between
the Euler number $\chi$ of the eight-manifold and the 
Pontryagin numbers,  $p_1$ and 
$p_2$ 

\eqn\aax{p_1^2-4p_2+8 \chi =0. }
The Pontryagin numbers are obtained by integrating the first and second
Pontryagin forms {\agw}
\eqn\aaax{ P_1=-{1\o 2} {\rm tr} R^2\qquad {\rm and} \qquad 
P_2=-{1\o 4} {\rm tr} R^4+{1\o 8} ({\rm tr} R^2)^2, } 
over $K^8$.
Replacing the spinor field in the 
$8_c$ representation by a spinor in the $8_s$ representation of 
$SO(8)$ corresponds to a change of sign in {\aax} 
\eqn\aaxiix{p_1^2-4p_2-8 \chi =0.} 
Therefore if one asks for an $8_c$ and an $8_s$ 
nowhere-vanishing spinor field,
one concludes that the Euler number of $K^8$ has to vanish {\ipw}.
However, it is also true that for every manifold having
$-8\chi=p_1^2-4p_2$ we can find another one which has  $8\chi=p_1^2-4p_2$,
obtained by reversing the orientation of the original manifold.
This corresponds to interchanging positive and negative
chirality spinors.

Comparing {\avii} with {\aaax} we observe that the anomaly polynomial
${X}_8$ is proportional to $P_1^2-4P_2$
and is therefore related to the Euler number of $K^8$
\eqn\aaix{ \int_{K^8} {X}_8=-{1\o 4! (2\pi)^4} \chi , }
which is a topological invariant.
Finding nowhere-vanishing spinors of both chiralities as
a solution of {\div} thus implies that the integral of the 
anomaly polynomial {\aaix} vanishes. 
Compactifications of eleven-dimensional supergravity on eight-manifolds
of this type 
have been considered in 
\ref\pepo{C.~N.~Pope and P.~van~Nieuwenhuizen, ``Compactifications 
of $d=11$ Supergravity on K\" ahler Manifolds'', \cmp {122} {1989} {281}.}.
For these compactifications no warp factor has been taken into account 
and the internal manifold is of the form $K^2\times K^6$, 
where $K^2$ is a two-dimensional 
sphere or torus and $K^6$ is a six-dimensional Calabi-Yau manifold. 
They yield
non-vanishing expectation values for the four-form field strength 
if the external space is anti-de Sitter and have an $N=4$ supersymmetry 
in three dimensions. 
However, we shall see that the situation is rather different 
if the anomaly is taken into account. In this case we will find 
solutions that preserve an $N=2$ supersymmetry if the external space
is three-dimensional Minkowski space, while the four-form 
field strength gets
a non-vanishing expectation value.

In compactifications with maximally symmetric three-dimensional 
space-time the non-vanishing components of $F_4$ are 
\eqn\bxii{\eqalign{ 
&F_{mnpq} \quad {\rm arbitrary,} \cr 
& F_{\mu \nu \r m} =\e_{\m\n\r} f_m, \cr}}
where $f_m$ is an arbitrary function that we will determine 
later on, as well as the explicit form of $F_{mnpq}$. 
$\e_{\m\n\r}$ is the 
completely anti-symmetric Levi-Civita tensor of $M^3$.  

Consider now the $\m$-component of the gravitino transformation 
law. Using {\div}, {\axi} and {\bxii} we obtain
\eqn\bxiii{
\eqalign{ 
\d \psi_{\mu} ={\nabla}_{\m} \eta & - {  1\o 288}  \D^{3/2}
(\g_{\m} \otimes \g_9 \g^{mnpq} ) F_{mnpq} \eta \cr 
& +{ 1\o 6}\D^{3/2} ( \g_{\m} \otimes \g^m ) f_m \eta \cr 
& -{1\o 4}\partial_n (\log \D)  
( \g_{\m} \otimes \g_9 \g^n )  \eta . \cr} } 
The simplest way to satisfy the condition $\d \psi_{\mu}=0$ is 
to consider compactifications of ${\cal M}$-theory 
to three-dimensional Minkowski space, so that we can find a spinor 
that satisfies: 
\eqn\namu{\nabla_{\mu}\e =0. }
Since we assume the three-dimensional space to be  
maximally symmetric, the above condition implies that 
the external space is Minkowski.

Using {\namu} we get that {\bxiii} can be satisfied if we set 
\eqn\axv{  F_{mnpq} \g^{mnpq} \xi =0, }
\eqn\biii{f_n=\partial_n \D^{-3/2}.} 
The second equation gives the explicit solution for one
 of the non-vanishing components of $F_4$
\eqn\baiii{F_{\m \n \r m}=\e _{\m \n \r}\partial_m \D^{-3/2}, }
in terms of the warp factor.

Next we consider the $m$-component of the gravitino 
transformation law. Using the properties of the gamma-matrices of the 
appendix and equations {\div}, {\bxii}, {\axv} and {\biii} we obtain: 
\eqn\axvi{ \d \psi_m = \nabla_m \x 
+{1\o 24}\D^{3/2} \g^{npq} F_{mnpq} \x + 
{1\o 4}  {\partial}_m (\log \D) \x 
-{3\o 8} \partial_n (\log \D) {\g_m}^n \x.  } 
It is now convenient to introduce transformed quantities: 
\eqn\dviii{\eqalign{
& {\tilde g}_{mn} =\D^{-3/2} g_{mn} , \cr 
& {\tilde \x} =\D^{1/4} \x, \cr}}
in terms of which the condition {\aaiixx} takes the simple form 
\eqn\dix{{\tilde \nabla}_m {\tilde \x} +{1\o 24} \D^{-3/4}
 F_m {\tilde \x}=0,}
where we have introduced the notation $F_m={\tilde \g}^{npq} F_{mnpq}$.
The relation {\dix} guarantees the existence of a covariantly constant
spinor
\eqn\dxi{{\tilde \nabla}_m ( {\tilde \x}^{\dagger} 
{\tilde \x}) =0, }
and its norm can be chosen to be one 
\eqn\axx{ 
{\tilde \x}^{\dagger} {\tilde \x} =1. }
For the components of $\tilde \xi=\tilde \xi_1+i \tilde \xi_2$ we can choose 
\eqn\bxxii{\eqalign{& \tilde \xi_i^T \tilde \xi_i ={1\o 2} 
\qquad {\rm for} \qquad i=1,2 \cr 
& \tilde \xi_i^T \tilde \xi_j =0 \qquad {\rm for} \qquad i\neq j. }}

Now we would like to show that $K^8$ is a complex manifold.  
In terms of ${\tilde \xi}$, we can construct an almost complex 
structure
\eqn\axxii{
{\tilde J}_m^{\;\;\; n} =i{\tilde \x}^{\dagger} {\tilde \g}_m^{\;\;\; n}
{\tilde \x}, } 
which is covariantly constant 
\eqn\axxiii{{\tilde \nabla}_p {\tilde J}_m^{\;\;\; n}=0. }
This can be easily seen taking into account that ${\tilde \x}^{\dagger}
{\tilde \g ^{a_1\dots a_n}}{\tilde \x}$ vanishes if $n$ is odd,
since the spinor involved is Weyl and $\g_9$ can be pulled 
through the expression.
The tensor {\axxii} has the property: 
\eqn\axxiv{ {\tilde J}_m^{\;\;\; n} {\tilde J}_n^{\;\;\; p}= 
-{\tilde \d_m}^{\;\;\; p} . }
To do this computation
it is convenient to use some formulas appearing in
\ref\cdfn{E.~Corrigan, C.~Devchand, D.~B.~Fairlie 
and J.~Nuyts, ``First-Order Equations for 
Gauge Fields in Spaces of Dimension Greater than 
Four'', \np {214} {1983} {452}.}
\ref\strohar{J.~A.~Harvey and A.~Strominger, ``Octonionic 
Superstring Solitons'', \prl {66} {1991} {549}.} which are expressed 
in terms of a fourth-rank antisymmetric tensor
\eqn\yi{\O_i^{mnpq}={\tilde \xi}_i ^T {\tilde \g}^{mnpq} 
{\tilde \xi}_i 
\qquad {\rm for} \qquad i=1,2. }
With the above normalization for the spinors it follows from 
{\cdfn}{\strohar} 
\eqn\yii{\O_i^{mnpq} \O_{i\; mnpq}=84. }
Furthermore using the Fierz rearrangement we can show that 
this tensor satisfies: 
\eqn\yiii{{\O_i}^{mnpq}\O_{j\; mnpq} =-12 \qquad {\rm for } \qquad 
i\neq j.} 

${\tilde J}_m^{\;\;\;n}$ is a complex structure since 
the Nijenhuis tensor 
\eqn\axxv{
{N_{mn}}^p={\tilde J}_m^{\;\;\; q} {{{\tilde J}_{[n;q]}}}^{\;\;\;  p}-
{\tilde J}_n^{\;\;\; q} {{{\tilde J}_{[m;q]}}}^{\;\;\; p}, } 
vanishes. Equation {\axxiv} together 
with {\axxv} imply that  
$K^8$ is a complex manifold. We are then allowed to 
introduce complex coordinates as well as holomorphic and anti-holomorphic 
indices, which we will denote with $a,b,\dots$ and ${\bar a}, 
{\bar b}, \dots$
respectively.
The metric $\tilde g_{mn}$ is of type $(1,1)$ and it is related to 
the complex structure as follows
\eqn\cxxia{\tilde J_{a\bar b}=i {\tilde g_{a\bar b}}.}
Since ${\tilde J}_{a \bar b}$ is covariantly constant, according to 
{\axxiii},  
it follows that $K^8$ is K\"ahler and ${\tilde J}_{a\bar b}$ 
is the K\"ahler form.
From equation {\cxxia} it follows that $\tilde \g_{\bar a}$
and $\tilde \g^a$  act as annihilation 
operators 
\eqn\cai{\tilde \g_{\bar a} {\tilde \x}={\tilde 
\g^a{\tilde \x}=0}.}

Next we would like to obtain the explicit form of the 
solution for the four-form field strength. Multiplying {\dix} with 
${\tilde \g}^a $ and using {\cai} we obtain the condition 
\eqn\caiii{F_{mnpq} {\tilde \g}^a {\tilde \g}^{npq} {\tilde \xi}=0, }
which is more restrictive than {\axv} and will allow us to obtain
the solution for $F_{mnpq}$.
All the components of the above expression must
vanish separately. 
From the equation
\eqn\caiiixx{F_{abcd} {\tilde \g}^{abcd} {\tilde \xi}=0,} 
we obtain the solution
\eqn\caiia{F_{abcd}=0. }
This can be easily seen by using {\cai} and the identity
\eqn\civ{F_{abcd}={1\o 384} F_{efgh} {\tilde \xi}^ {\dagger} 
\{{\tilde \g}_{abcd} , 
{\tilde \g} ^{efgh}\} {\tilde \xi} ,}
which follows from properties of gamma matrices of the appendix.
By complex conjugation of {\caiia} we get
\eqn\caiiaiix{
F_{{\bar a}{\bar b}{\bar c}{\bar d}}=0. }
Similarly one gets from the equation 
\eqn\hi{F_{a {\bar b} {\bar c} {\bar d}} {\tilde \g}^e 
{\tilde \g}^{{\bar b} {\bar c} {\bar d}} \xi=0, }
the result 
\eqn\daiia{F_{ a{\bar b}{\bar c}{\bar d}}=0.}
By complex conjugation it follows 
\eqn\ddaiia{F_{{\bar a} b c d}=0. }
The vanishing of the remaining components of {\caiii} can be written
in the form
\eqn\caiiia{F_{{\bar a} b {\bar c} d}{\tilde J}^{{\bar c}  d}=0. }
This expression reminds the Donaldson-Uhlenbeck-Yau equation
\ref\gsw{M.~Green, J.~H.~Schwarz and E.~Witten, ``Superstring Theory'', 
Cambridge Univ. Pr. (1987). } appearing in the heterotic 
string {\strom}. However, in this case the field strength is a 
four-index object instead of a two-index object and the ``gauge group'' 
is abelian instead of $SU(N)$.  

It is satisfying to see that equations {\caiia},
{\caiiaiix}, {\daiia}, {\ddaiia}
 and {\caiiia} 
represent a solution of the field equation {\aixxi}. 
In fact, since we have derived these results from supersymmetry it is 
natural to think that they will solve the field equation for $F_4$, 
which takes the form:
\eqn\ddi{{\hat E}^{-1} \partial_m 
\left( {\hat E} {\hat F}^{mnpq}\right)  ={1\o 16} 
{\hat \e}^{npqrstuv} \D^{-1} \partial_{r} \D {\hat F}_{stuv}. }
Here we have used that {\aviii} is conformally invariant,
so that the contribution of $X_8$ to the field equation {\aixxi} 
vanishes for this component of $F_4$.
Choosing a basis in which the metric is diagonal and 
using the explicit form of the 
four-form field strength this equation can be transformed to 
\eqn\ddii{\partial_{[{\bar a}} F_{b {\bar c} d{\bar e}]} =0, }
which is nothing but the Bianchi identity {\aaiaai}.
To summarize, the only non-vanishing components of $F_4$ are 
$F_{\m \n \r m}= \e_{\m \n \r} f_m$ and $F_{a {\bar b} c {\bar d}}$.

Taking this result into account and {\caiiia} it is 
easy to see that from expression {\dix} we get
\eqn\cxaii{{\tilde \nabla}_m {\tilde \x}=0,}
so that $K^8$ is Ricci flat. From this equation 
it follows ${\tilde R}_{mn}=0$. 
Since we already showed that $K^8$ is K\"ahler, 
we conclude that the holonomy group is $SU(4)$ and that the internal
manifold is a Calabi-Yau four-fold. These manifolds 
have vanishing first Chern class.  
The original metric appearing in {\dviii} is
not K\"ahler but conformal to the K\"ahler metric.
These metrics
are called ``conformally Calabi-Yau'' {\strom}.
It is useful to recall at this point some properties of 
Calabi-Yau four-folds.
Since the holonomy group is $SU(4)$
there are two covariantly constant spinors 
for a given chirality that come from
the decomposition $8_c\rightarrow 6 \oplus 1 \oplus 1$ 
under the reduction 
of $SO(8)$ to $SU(4)$.
The two singlets of this decomposition correspond to the 
two real covariantly 
constant spinors for a given chirality that we found.

We still need to find the explicit form of the solution 
to equation {\caiiia}
on a K\"ahler space. 
Roughly as in {\strom} or
\ref\post{J.~Polchinski and A.~Strominger, 
``New Vacua for Type II String Theory'', preprint UCSBTH-95-30, 
hep-th/9510227. }, $F_{a \bar b c \bar d}$ can be written in terms 
of the harmonic four-forms $\omega_{(4)}^i$ on $K^8$ 

\eqn\bii{F=\sum_{i=1}^{h_{11}}\nu^{(4)}_i \omega_{(4)}^i.  } 
Here the $\n ^i$'s are constants, $h_{11}$ are the Hodge numbers and
$F$ is a shorthand notation for the above component of $F_4$.
There are several constraints on the $\n ^i$'s. The quantization of the 
magnetic charge implies that these constants should be integers. 

A second constraint can be obtained from the Bianchi
identity as follows.
Inserting the solution {\biii} into the Bianchi identity {\aixxi} 
gives an equation for the warp factor
\eqn\biv{d*d \log \Delta={1\o 3} F\wedge F -
{2\o 3} (2\pi)^4 X_8  . }
Therefore, integrating the Bianchi identity over the eight-manifold 
we obtain a relation between the characteristic class 
represented by $\int F \wedge F$ and the Euler number  
\eqn\aaaxii{\int_{K^8} F \wedge F +{1\o 12} \chi =0, }
where we have used Stoke's theorem and we have imposed 
the condition that $*F_4$ should be globally defined.  
For complex manifolds of real 
dimension eight there is a relation between the Euler number 
and the $4^{\rm th}$ Chern class $\chi=c_4(M)$ so that 
{\aaaxii} can be written in the form
\eqn\aaaxiii{\int_{K^8} F \wedge F+{1\o 12} c_4(M)=0. }
Inserting {\bii} into {\aaaxiii} we obtain 
conditions on the constants $\nu_i^{(4)}$. Furthermore,  
the constraint {\aaaxiii} provides a topological restriction on the possible 
compactifications of ${\cal M}$-theory to three dimensions.

At this point we have determined a complete solution 
to the supersymmetry transformations. 
Equation {\axxiii} and {\cxaii} state that 
the internal manifold is a Calabi-Yau four-fold, while 
the original metric {\dviii} is non-K\"ahler but conformal to a K\"ahler 
metric. One of the non-vanishing components of $F_4$
has to satisfy {\caiiia} and can be expressed in terms 
of harmonic four-forms {\bii}. The coefficients of this expansion should
obey the constraint {\aaaxiii}.
Equation {\biv} is an 
equation for the warp factor. Finally, 
the warp factor determines the remaining 
non-vanishing component of $F_4$ {\baiii}. 

\newsec{Conclusion and Outlook}
We have shown the existence of new vacua of ${\cal M}$-theory
compactified on an eight-manifold that preserve an $N=2$ supersymmetry
in $D=3$. For these compactifications a warp factor for the metric
has been taken into account, which is non-trivial in three 
space-time dimensions. 
Due to this fact, the four-form 
field strength acquires a non-vanishing expectation value
for compactifications to three-dimensional Minkowski space-time.
This is surprising and in constrast to the situation 
appearing in conventional
compactifications of eleven-dimensional supergravity, where 
the expectation value of the four-form field strength has
to vanish, if supersymmetry is unbroken
\ref\cara{P.~Candelas and D.~J.~Raine, ``Compactification 
and Supersymmetry in $d=11$ Supergravity'', \np {248} {1984} {415}.}.

While the original metric on the internal space is not K\"ahler, 
it can be conformally transformed to a metric that is K\"ahler and
Ricci flat, so that the internal manifold has $SU(4)$ holonomy.

This, of course, implies the existence of new vacua
for the type IIA string theory in two dimensions.
A crucial ingredient to get this result was the existence
of the anomaly in the eleven-dimensional supermembrane action,
which appears as a consequence of membrane-fivebrane duality.
Such an anomaly is not present in the type IIB string theory,
and a similar computation for compactifications on four- 
and six-manifolds gives a constant warp factor, and vanishing
expectation values for the field strength for compactifications 
to Minkowski space {\wshd}
\ref\bbsu{K.~Becker, M.~Becker and A.~Strominger, unpublished.}.

We have considered manifolds for which the holonomy group is $SU(4)$. 
This leads, as we have explained, to two covariantly 
constant spinors of a well defined chirality. One could also 
consider compactifications on manifolds that admit only one 
covariantly constant spinor. Examples of these manifolds 
are 8-manifolds with Spin(7) holonomy. 
The corresponding spinor arises from the 
decomposition $8_c\rightarrow 7 \oplus 1$ {\ipw}. 
Compactifications on these manifolds have attracted recently 
some attention in connection to ${\cal F}$-Theory, 
and it would be interesting 
to carry out a similar computation as the one presented herein.

\vskip 1cm
\noindent {\bf Acknowledgements}

\noindent We are grateful to M.~Duff, J.~Harvey, 
D.~Lowe, H.~Nicolai, J.~Polchinski, R.~Schimmrigk, 
S.-T.~Yau and very specially A.~Strominger for useful discussions. 
This work was supported by
DOE grant DOE-91ER4061  and NSF grant PHY89-04035.

\vskip 1cm 
\noindent {\bf Appendix}

\no Our notation and conventions are as follows
\item{\tria} The different types of indices that we use are: 
\eqn\indices{\eqalign{
M, \, N, \dots \qquad &\hbox{ are eleven-dimensional 
world indices},\cr 
A, \, B, \dots \qquad &\hbox{ are eleven-dimensional 
tangent space indices},\cr
m,\; n,\dots \qquad &\hbox{are real indices of the 
euclidean submanifold},\cr
a, \; b, \dots 
{\rm and} \; {\bar a}, \; {\bar b}, \dots \qquad &
\hbox{are complex indices of the euclidean submanifold},\cr 
\m, \; \nu, \dots \qquad 
&\hbox{are three-dimensional lorentzian indices},\cr} }
We denote by $(x^0,x^1,x^2)$ the coordinates of the external space, 
while $(x^3, \dots, x^{10})$ are the coordinates of the 
eight-manifold. 

\item{\tria} $\e^{MNPQRSTUVWX}$ denotes a tensor, rather than a 
tensor density, with 
\eqn\lete{\e^{012\dots 10}={\hat E}, }
and analogously for the Levi-Civita tensors of $M^3$ and 
$K^8$ respectively. 

\item{\tria} $n$-forms are defined with a $1/n!$. For example: 
\eqn\form{F= {1\o 4!} F_{mnpq} dx^m \wedge dx^n \wedge dx^ p 
\wedge dx^q.}
\item{\tria} The gamma-matrices ${\hat \G}^M$ are hermitian, 
for $M=1, \dots, 10$ while ${\hat \G}_0$ is antihermitian. 
They satisfy: 
\eqn\algebra{\{ {\hat \G}_M, {\hat \G} _N \} =2{\hat g}_{MN}, }
where ${\hat g}_{MN}$ has the signature $(-,+,\dots,+)$. 
${\hat \G}_{M_1 \dots M_n}$ is the antisymmetriced product of gamma 
mat  ices: 
\eqn\antis{{\hat \G}_{M_1 \dots M_n}= {\hat \G}_{[M_1} \dots 
{\hat \G}_{M_n]},  }
where the square bracket implies a sum over $n!$ terms with a $1/n!$ 
prefactor.
\item{\tria}A representation of the $d=3$ gamma matrices is 
\eqn\threega{
\g_0=\left( \matrix{0 & i \cr  i &0\cr }\right),  
\qquad \g_1=\left( \matrix{ 0 &-i \cr i & 0 \cr }\right),  \quad 
{\rm and} \quad  
\g_2=\left( \matrix{ 1 & 0 \cr  0 & -1 \cr} \right), }
they satisfy $\e_{\m\n\r} \g^{\n\r} =2\g_{\m}$. 
\item{\tria} Gamma matrix identities that are useful are
\eqn\cuga{[\nabla_m, \nabla_n]\xi 
={1\o 4} R_{mnpq} \g^{pq} \xi, }
\eqn\gaii{\eqalign{ & [\g_m, \g^r]=2{\g_m}^r, \cr 
& \{\g_{mn}, \g^r \} = {2\g_{mn}}^r, \cr 
& [\g_{mnp}, \g^r ]=2{ \g_{mnp}}^r,\cr 
& \{ \g_{mnpq}, \g^r \} =2 
{\g_{mnpq}}^r,\cr }\qquad  
\eqalign{& \{ \g_m, \g^r\}=2{\d_m}^r , \cr
&[\g_{mn} ,\g^r ]=-4{\d^r}_{[m} \g_{n]},\cr 
& \{ \g_{mnp}, \g^r\} =6 {\d^r}_{[m} \g_{np]}, \cr   
& [\g_{mnpq}, \g^r] =-8{\d^r}_{[m} \g_{npq]}.\cr }  }
\item{\tria}The chirality operator is defined by 
\eqn\chiral{\g_9 = {1\o 8!}{\e}_{mnpqrstu} {\g}^{mnpqrstu}.}
\item{\tria}We use the Fierz identity 
\eqn\fierz{\chi {\bar \psi} ={1\o 2^{[d/2]}}  \sum_{n=0}^d 
{1\o n!} \G^{c_n \dots c_1} {\bar \psi} \G_{c_1 \dots c_n} \chi .}
\item{\tria}Our definition of Hodge $*$ is: 
\eqn\hodge{*(dx^{m_1} \wedge \dots \wedge dx^{m_p})  
={1\o (d-p)!}{\e^{m_1 \dots m_p}}_{m_{p+1} \dots m_d} 
dx^{m_{p+1}} \dots \wedge dx^{m_d}. }

\listrefs
\end